\documentstyle[11pt,newpasp,twoside]{article}
\newcommand{\gapr}{\raisebox{-.6ex}{\mbox{
$\stackrel{>}{\mbox{\scriptsize$\sim$}}\:$}}}

\reversemarginpar

\begin{document}
\title{
The Puzzling Compact Objects in Supernova Remnants}
 \author{G.\ G.\ Pavlov, D.\ Sanwal, G.\ P.\ Garmire}
\affil{Pennsylvania State University, Department of Astronomy \&
Astrophysics, 525 Davey Lab, University Park, PA 16802}
\author{V.\ E.\ Zavlin}
\affil{Max-Planck-Insitut f\"ur Extraterrestrische Physik, Garching, Germany}

\begin{abstract}
X-ray images of some young supernova remnants show bright point sources 
which have not been detected in radio, optical
and gamma-ray bands. Despite the similarity of the X-ray spectra of these 
objects, they show a variety of temporal properties. Most
likely, they are neutron stars whose properties 
(spin periods? magnetic fields?
environments?) are different from those of radio and/or 
gamma-ray pulsars. 
We present an overview of observational results
on several objects
of this class --- the central sources of Cassiopeia A, RX J0852--4622, RCW 103,
Puppis A, and PKS 1209--51/52 ---
with emphasis on the recent {\sl Chandra} observations.
\end{abstract}
%%%%%%%%%%%%%%%%%%%%%%%%%%%%%%%%%%%%
\section{Introduction}
There are two types of isolated (non-binary) neutron stars (INSs)
with quite different observational manifestations.
{\it Active, rotation-powered pulsars}
include, in addition to commonly known radio pulsars,
some ``radio-quiet'' 
INSs whose radio pulsations are not seen because of
 unfavorably directed pulsar beams.
Pulsar activity in these radio-quiet active pulsars
 can be seen at wavelengths other than radio
(e.g., in $\gamma$-rays --- the classical example is Geminga), 
and it can manifest itself, in young
pulsars, through a compact pulsar-wind nebula (PWN).
A distinctive feature of active pulsars is that these objects are
powerful sources of {\em nonthermal radiation from NS magnetospheres},
in radio through $\gamma$-rays. 
The intensity of this nonthermal
radiation usually exceeds that of the surface
(thermal) radiation, at least in young pulsars.

Recent observations have established that
there exist isolated compact objects,
presumably INSs, which show no signs of pulsar activity ---
i.e., the usual pulsar processes, which
result in beams of relativistic particles, PWNe, and
strong nonthermal radiation, do not operate in these objects.
They are often called {\it radio-silent INSs} (or INS candidates),
{\it failed pulsars}, {\it dead pulsars}, etc ---
none of these terms describes their properties adequately
(e.g., some of these objects do show X-ray pulsations,
but their origin is apparently quite different from that
observed from `ordinary' pulsars). 
These objects can be further divided in at least four classes, based
on their observational manifestations. First, 
six of them
have been dubbed {\it Anomalous X-ray Pulsars} (AXPs;  
Mereghetti \& Stella 1995; van Paradijs, Taam, \& van den Heuvel 1995)
because they resemble accreting binary pulsars, but their luminosities,
$L_{\rm x}\sim 10^{35}$ erg s$^{-1}$,
are two-three orders of magnitude lower, their spectra are considerably
softer than those of binary pulsars, 
and their periods cluster in a narrow range of 6--12 s
(Israel, Mereghetti, \& Stella 2001). The second class is comprised of
three (perhaps, four) {\it Soft Gamma-ray
Repeaters} (SGRs; Kouveliotou 1995)
whose quiescent
radiation shows many similarities with AXPs, but, contrary to
AXPs, SGRs occasionally undergo strong outbursts,
with peak luminosities up to $10^{44}$ erg s$^{-1}$.
Thompson \& Duncan (1996) suggested that AXPs and
SGRs are {\it magnetars} --- INSs with superstrong magnetic fields,
$\sim 10^{14}$--$10^{15}$ G. 
Some 
of the AXPs and SGRs are apparently associated with 
supernova remnants (SNRs).
Recent results on AXPs and SGRs are
presented by S.\ Kulkarni (this volume).
One more class of presumably older 
``{\it truly isolated}'' 
(i.e., not associated with SNRs), radio-silent NSs has emerged recently 
(e.g., Treves et al.\ 2000).
These objects show very soft, thermal-like 
X-ray radiation with temperatures 
$kT\sim 50$--150 eV
and luminosities $\sim 10^{32}$--$10^{33}$ erg s$^{-1}$.
Finally, {\it Compact Central Objects} (CCOs)
have been found in several SNRs,
which have not been identified as active pulsars,  AXPs or SGRs.
These objects
are particularly puzzling. 
Below we will discuss recent results on five CCOs
which have been  observed with the {\sl Chandra} X-ray Observatory.
%%%%%%%%%%%%%%%%%%%%%%%%%%%%%%%%
\section{Observational Results}
%%%%%%%%%%%%%%%%%%
\subsection{General Overview}
%%%%%%%%%%%%%%%%%%
Observational properties of CCOs are summarized in Tables 1 through 5.
First column of 
Table 1 gives the positions for the four objects observed
with {\sl Chandra} in an imaging mode (accurate to $1''$--$2''$)
 and the most accurate {\sl ROSAT}
HRI position (uncertainty $\sim 5''$)
for the fifth source. 
In the other columns, the host SNRs, 
with their ages and distances, are listed. The references here and in the
other tables are given only to recent works, where the references to
earlier results can be found.
%%%%%%%%%%
  \begin{table*}
  \begin{center}
  \begin{tabular}{ccccc}
\tableline
 Object &  SNR & Age & $d$ & Ref.\\ 
 {} & {} & kyr & kpc &  \\
\tableline
CXO~J232327.9+584843  & Cas A       & 0.32       & 3.4(3.3--3.7) & 1 \\
CXO~J085201.4--461753 & G266.1--1.2 & $\sim$1--3 & 1(1--2)       & 2 \\
CXO~J161736.3--510225 & RCW~103     & 1--3       & 3.3(3--7)      & 3 \\
CXO~J082157.5--430017 & Pup A       & 1--3       & 2.2(1.6--3.3) & 4 \\
RX~~J121000.8--522625 & G296.5+10.0 & 3--20      & 2.1(1.3--3.9) & 5 \\
\tableline\tableline
\end{tabular}
\caption{Compact Central Objects in SNRs.
\footnotesize{
References: 1- Murray et al.\ (2001);
2- Pavlov et al.\ (2001);
3- Garmire et al.\ (2000a), Leibowitz \& Danziger (1983);  
4- Zavlin et al.\ (1999), this work; 5- Zavlin et al.\ (1998).}}
\end{center}
\end{table*}

Table 2 gives the periods $P$ (detected or suspected
--- the latters are marked with `[?]'),
 the observed 
X-ray fluxes $F_{\rm x}$ (in units of $10^{-12}$ erg cm$^{-2}$ s$^{-1}$)
in 
energy
bands
$\Delta E$, and limiting optical/IR magnitudes 
for the same five CCOs (abbreviated positions in first column).
We see that the CCOs are X-ray bright --- in fact, three of them,
the CCOs of RCW 103, Pup A and G296.5+10.0, 
were discovered with
{\sl Einstein}, and the Cas A CCO was found in the archival {\sl Einstein}
data.
On the other hand, no optical, radio or $\gamma$-ray
  counterparts have been identified
for these objects, except perhaps the RCW 103 CCO (see below).
The fluxes of four CCOs have not shown long-term variabilities,
but the RCW 103 CCO is clearly variable (Gotthelf, Petre, \& Vasisht 1999)
--- we present the lowest and highest flux values observed with
{\sl Chandra}.
%%%%%%%%%%%%%%%%%%%%%%
\begin{table*}
\begin{center}
\begin{tabular}{ccccccc}
\tableline
Object & $P$ & $F_{\rm x,-12}$ & $\Delta E$ & Optical/IR & Ref. \\
       &     &                 & keV        & mag        &  \\
\tableline
J2323+5848  & 12ms[?]& 0.8     & 0.3--6  & R$>$27.8, m$_{675}$$>$28.9 & 1 \\
            &        &       & & J$>$22.5, K$_s$$>$21.2 &  \\
J0852--4617 & ...    & 1.4     & 0.4--6  & B$>$22.5, R$>$21.0 &  2 \\
J1617--5102 & 6hr[?] & 0.9--60 & 0.5--9 & J$\approx$22.3, H$\approx$19.6, K$_{s}$$\approx$18.5 [?] &  3
 \\
J0821--4300 & ...  & 4.5     & 0.4--6  & $r^\prime$$>$24.8, $i^\prime$$>24.4$, $z^\prime$$>$22.4  &  4 \\
J1210--5226 & 424ms  & 1.9   & 0.5--6  & $r^\prime$$>$24.8, $i^\prime$$>24.4$, $z^\prime$$>$22.9 & 5 \\
\tableline\tableline
\end{tabular}
\caption{Periods, X-ray fluxes, and optical/IR magnitudes.
\footnotesize{
References: 1- Murray et al.\ (2001), Pavlov et al.\ (2000), 
Kaplan et al.\ (2001),
Fesen et al.\ (this volume); 
2- Pavlov et al.\ (2001), this work;
3- Garmire et al.\ (2000b), this work, 
Sanwal et al.\ (2002); 
4- this work, Wang \& Chakrabarty (this volume);
5- Zavlin et al.\ (2000), this work, Wang \& Chakrabarty (this volume).}} 
\end{center}
\end{table*}

Table 3 gives the results of fits
of the {\sl Chandra} ACIS spectra with the `standard' blackbody (BB)
and power-law (PL) models.
The bolometric luminosities $L_{\rm bol}$
 for the BB
fits, and the luminosities $L_{\rm x}$ for the PL
fits
in the 
bands
$\Delta E$ 
from
Table 2, are in units of 
$10^{33}$ erg s$^{-1}$. The luminosities and the effective radii $R$
are 
for most plausible distances (see Table 1).
The parameter $\gamma$ 
is the photon index. The hydrogen
column densities are in units of $10^{22}$ cm$^{-2}$.
%%%%%%%%%%%%%%%5
\begin{table*}
\begin{center}
\begin{tabular}{c|cccc|ccc|c}
\tableline
 Object & $kT$ & $R$ & $L_{\rm bol,33}$ & $n_{\rm H,22}$ & $\gamma$ &  $L_{\rm x,
33}$ & $n_{\rm H,22}$ & Ref.\\
   &  keV & km &
     & & & & & \\
\tableline
J2323+5848  & 0.49 & 0.5     & 1.6    & 1.1 & 4.1 & 500   & 2.2 & 1 \\
J0852--4617 & 0.40 & 0.3     & 0.3    & 0.4 & ... & ...   & ... & 2 \\
J1617--5102 & 0.4--0.6 &0.2--1.6 & 0.5--30 & 1.5 & 4--5 & 10--600 & 3.7 & 3 \\
J0821--4300 & 0.38 & 1.4 & 4.2 & 0.2 & 4.8 & 80 & 1.0 & 4 \\
J1210--5226 & 0.25 & 1.6  & 1.3 & 0.04 & 5.1 & 8   & 0.5  & 5  \\
\tableline\tableline
\end{tabular}
\caption{Best-fit fit parameters for blackbody and power-law fits.
\footnotesize{
References: 1- Murray et al.\ (2001); 2- this work; 3- Garmire et al.\ (2002);
4,5- this work.}}
\end{center}
\end{table*}
%%%%%%%%%%%%%%%%%%%%%%%%%%%%%%%%%%%%
In some cases, the `standard'
models do not fit the observed spectra well ---
 e.g., the recently observed {\sl Chandra} ACIS spectrum of the
G266.1--1.2 CCO is too steep at higher energies to fit a PL
model. 
The fits with the simple BB
model
are formally acceptable for most CCOs.
They give temperatures in a narrow
range of 0.3--0.6 keV, but the effective radii, 0.2--1.6 km,
are much smaller than the  expected NS radii, $R_{\rm NS}\approx 10$--15 km.
On the other hand, the PL
spectra (where they fit the data)
are very steep 
(in comparison with typical $\gamma \approx 1.5$--2
measured for active radio pulsars), and the PL
fits yield
too high values of $n_{\rm H}$. It is quite plausible that none
of these simplistic fits provide true physical parameters of
the sources. Therefore, we also present,
 in Table 4, the fits with the magnetic NS hydrogen
atmosphere models (Pavlov et al.\ 1995)
which generally give lower effective
temperatures and larger emitting areas than the BB
fits.
In the case of two of the five CCOs
(in Pup A and G296.5+10.0), the radii become close
to the expected NS radii, but they are still too small for the other
three CCOs. 
%%%%%%%%%%%%%
\begin{table*}
\begin{center}
\begin{tabular}{cccccc}
\tableline
Object & $kT_{\rm eff}^\infty$ & $R^\infty$ & $L_{\rm bol}^\infty$ & $n_{\rm H}$ & Ref. \\
  &  keV & km & $10^{33}$erg s$^{-1}$ & $10^{22}$cm$^{-2}$ & \\
\tableline
J2323+5848  & 0.3  & 1   & 2     & 0.8 & 1 \\
J0852--4617 & 0.27 & 1.5 & 0.8   & 0.5 & 2 \\
J1617--5102 & 0.3  & 1--8 & 1--60 &  1.7 & 3 \\
J0821--4300 & 0.17 & 10 & 8   & 0.4  & 4 \\
J1210--5226 & 0.14 & 11  & 1     & 0.2 & 5 \\
\tableline\tableline
\end{tabular}
\caption{Fits with magnetic hydrogen atmosphere models.
The subscript $^\infty$ denotes the values as seen by a distant observer.
\footnotesize{
References: 1- Pavlov et al.\ 2000; 2,3,4,5- this work.}}
\end{center}
\end{table*}

For comparison with the AXP and SGR quiescent spectra,
which are well described by a two-component BB+PL spectral model,
we also present, in Table 5, the PL+BB fits for three CCOs 
(the PL component is unconstrained for the two other CCOs).
As a rule, the additional PL component
(20\%--30\% of the observed energy flux) only slightly improves the 
quality of the fits. 
%%%%%%%%%%%%%
\begin{table*}
\begin{center}
\begin{tabular}{ccccccc}
\tableline
 Object & $kT$ & $R$ & $\gamma$ & $n_{\rm H,22}$ & $F_{\rm x}^{\rm bb}/F_{\rm x}^{\rm pl}$ & Ref. \\
   &  keV & km & & & &  \\
\tableline
J2323+5848 & 0.45  & 0.54 & 1.7   & [1.1] & 3.8 & 1 \\  
J1617--5102 & 0.52 & 1.5 & 3.8     & 2.2 & 2.2 & 3\\      
J1210--5226 & 0.22 & 2.0 & 3.6     & 0.13& 3.0 & 5\\
\tableline\tableline
\end{tabular}
\caption{Best-fit fit parameters for BB+PL fits. The
hydrogen column density was fixed in the J2323+5848 fit.
The fit for J1617--5102 is for 2001 Oct 7 data.
\footnotesize{
References: 1- Murray et al.\ (2001); 3,5- this work. }
}
\end{center}
\end{table*}
%%%%%%%%%%%%%%%%%%%%%%%%%
\subsection{Individual Sources}
%%%%%%%%%%%%%%%%%%%%%%%%%
%%%%%%%%%%%%%%%%%%%%%%%%%%%%%%%%%%%
\subsubsection{Cas A CCO:} 
The central source of Cas A,
the youngest Galactic SNR, was discovered in the first-light
{\sl Chandra} observation (Tananbaum 1999) and 
after that found in the archival {\sl ROSAT} and {\sl Einstein}
images (Aschenbach 1999; Pavlov \& Zavlin 1999). 
The {\sl Chandra}
observations of this source
were analyzed by Pavlov et al.\ (2000), Chakrabarty et al.\ (2001), and
Murray et al.\ (2001).
The source is about $7''$ from
the Cas A
expansion center (Thorstensen et al.\ 2001), which corresponds
to the transverse velocity of $\simeq 330$ km/s, for the age of 320 yrs
and distance of 3.4 kpc.
Deep 
searches in the radio (McLaughlin et al.\ 2001) and optical/IR
(Kaplan, Kulkarni, \& Murray 2001; 
Fesen, Chevalier, \& Holt, this volume) 
failed to find a counterpart. The very large
X-ray-to-optical flux ratios (e.g., $F_{\rm x}/F_{\rm R} >3000$),
together with the proximity to the SNR expansion center,
prove that the point source is the compact remnant
of the SN explosion. 
A lack of a PWN around the source, 
the steep X-ray
spectrum, and low 
luminosity
argue that it is not an active,
rotation-powered pulsar.
The very high
BB temperature
and the very small size
hint that the radiation emerges from small heated regions on the NS
surface. If even the NS surface were covered with a light-element
(H or He) atmosphere, 
the effective temperature would still be too high for
a cooling NS of this age, and the size would be smaller
than the NS radius.
The origin of the heated regions on the NS surface remains unclear.
In principle, they could be explained by a very strong magnetic
field localized in a small region close to the surface, which would
locally increase the surface temperature due to enhanced thermal
conduction along the magnetic field, or due to heat caused by field decay.
This hypothesis, however, has not been supported by direct calculations.
The temperature nonuniformity might be also associated with a nonuniform
chemical composition of the crust/atmosphere --- for instance, light-element
polar caps, formed by accretion onto the magnetic poles of the nascent NS,
would be hotter than the rest of the surface comprised of
heavy elements (e.g., iron) because the low-Z envelopes are more
efficient heat conductors. However, it is not clear whether such a gradient
of chemical composition could survive for a long time, or what would be
the mechanism supporting this gradient if there is no accretion at the
present stage.

If the CCO radiation is emitted from small heated regions on the
NS surface, 
then it would be 
natural to expect that the radiation is pulsed.
The deepest search for pulsation, undertaken by Murray et al.\ (2001),
resulted in a candidate period of about 12 ms, whose significance, however,
is rather low. Such a period would not be easy to explain. On one
hand, it strongly limits the magnetic field --- a NS rotating
with such a period should be an active pulsar
at $B\gapr 10^9$ G, but we see no manifestations
of pulsar activity. On the other hand, if the magnetic field is
low enough to put the NS under the `pulsar death line',
it would be very difficult to explain the origin of the 
heated regions.

One could assume that the observed X-ray emission is caused by
accretion onto a NS or a black hole (BH), with an accretion rate
$\dot{M}\sim 10^{12}$ g s$^{-1}$. Although this rate is very low
compared to that in accreting compact objects in binaries, it is,
nevertheless, too high to be explained by accretion from the circumstellar
medium because  a very high density of the ambient matter, $n\gapr 10^5
(v/300~{\rm km/s})^3$~cm$^{-3}$, would be required.
We cannot completely rule out the possibility of accretion from
a very underluminous (e.g., ${\rm M_R}>11$) secondary companion,
but it would be 
much dimmer than secondaries
in usual low-mass X-ray binaries.
Somewhat more plausible
source of accretion might be a `fossil disk', left-over after the
SN explosion (van Paradijs et al.\ 1995), but such a disk should be much
less luminous in the optical than those detected in X-ray binaries.
Generally, any accretion model seems to contradict to the lack of 
variability of the X-ray radiation. 

To conclude, there is no full understanding of the real nature of
the Cas A CCO. In our opinion, most likely it is an INS emitting thermal
radiation from small heated surface areas. Critical observations
would be a firm detection of X-ray pulsations and deep IR observations.
%%%%%%%%%%%%%%%%%%%%%%%%%%%%%%%%
\subsubsection{CCO in ``Vela Junior'':}
The shell-like SNR G266.1--1.2 (= RX~J0852--4622) has been discovered quite
recently with {\sl ROSAT}
 at the south-east corner of the Vela SNR (Aschenbach 1998).
Possible detection of the 1.156 MeV $\gamma$-ray
line of the  radioactive isotope $^{44}$Ti 
with the {\sl CGRO} COMPTEL (Iyudin et al.\ 1998) may imply a very young age,
$\sim 700$ yr, at a distance of $\sim 200$ pc,  
although further observations
of the SNR have shown that it is likely older and more distant
(Slane et al.\ 2001).
There were a few reports about possible point-like X-ray sources close
to the SNR center (Aschenbach 1998; Slane et al.\ 2001; Mereghetti
2001), but it had been hard to rule out the association
of the CCO candidates with two bright stars in the field.
A 3-ks observation with the {\sl Chandra} ACIS allowed us
(Pavlov et al.\ 2001; Kiziltan et al., this volume) to
firmly detect the 
point source 
and measure its
position with high accuracy (see Table 1). 
A follow-up optical observation showed a star (B=18.9, R=16.9)
at $2\farcs4$ from the center of the {\sl Chandra} error circle,
but the association of this star with the CCO looks unlikely.
The limiting magnitudes of an optical counterpart (Table 2) are
not as deep as for the other CCOs, but the X-ray-to-optical flux
ratio is high enough to conclude that this is indeed the compact
remnant of the SN explosion, at $\approx 4'$ from the SNR center.
A follow-up 30-ks
 {\sl Chandra} ACIS observation 
allowed us to measure the source flux and the spectrum.
Interestingly, the spectrum does not fit the PL model,
but it fits the BB model and shows a temperature slightly below
that of the Cas A CCO. The size of the emitting area and the luminosity
cannot be accurately estimated because of the poorly known distance
(the values in Table 3 are normalized to $d=1$ kpc);
at a plausible distance of 2 kpc the size and the luminosity are close
to those of the Cas A CCO.
These similarities suggest that the
Cas A and Vela Junior CCOs indeed belong to the same family.

The second {\sl Chandra} ACIS observation
was carried out in Continuous Clocking (CC) mode,
which allowed us to search for pulsations.
A candidate period, $P=301$ ms, was found at a $2.8\,\sigma$ level
--- too low significance to claim it is real until it is confirmed
by another observation.
%%%%%%%%%%%%%%%%%%%%%%%%%%%%%%%%%%%%
\subsubsection{RCW 103 CCO:} 
Historically, it was the first radio-quiet, X-ray-bright NS
candidate found in a young SNR (Tuohy \& Garmire 1980). It has been studied with
{\sl Einstein} (Tuohy et al.\ 1983), {\sl ASCA} (Gotthelf, Petre, \&
Hwang 1997) and {\sl Chandra}
(Garmire et al.\ 1999, 2000ab). Contrary to the other CCOs, strong
(up to a factor of 70) long-term variations
of its flux have been found.
First
16-ks {\sl Chandra} ACIS observation (1999 September 26)
of this source showed a light curve which looked like a fraction
of a sinusoid with a period of about 6 hr; a similar period was apparently
found in the archival {\sl ASCA} data (Garmire et al.\ 2000a). In second
23-ks {\sl Chandra} ACIS observation of 2000 February 8 (Garmire et al.\ 2000b), 
the source was so bright
that it was difficult to measure its properties because of strong
pile-up (saturation) of the source image.  The source was certainly variable,
at a 20\% level, during
this observation, but the variability did not look periodic.
Further long X-ray observations will show whether the 6-hr period
is observable in the low state only.
A series of six short, 3--4 ks, {\sl Chandra} ACIS observations of the CCO 
in January--October 2001
showed that the source was in high state during this period
(flux around $1\times 10^{-11}$ erg cm$^{-2}$ s$^{-1}$ in the 0.5--9 keV band).
In most of the {\sl Chandra} observations, the source spectrum resembled
a blackbody; its temperature $kT\approx 0.4$--0.6 keV did not show
correlations with the flux, while the effective emitting area 
was approximately proportional to the flux, with a maximal size
of $\approx 1.6$ km in the high state. 
Fits with the hydrogen NS atmosphere models give, as usual, lower
temperatures and larger sizes, but even the increased effective radius
remains smaller than $R_{\rm NS}$, at least in low state. Adding a PL
component slightly improved the fits for some of the observations,
but it was poorly constrained in most cases.

The last of the short {\sl Chandra} ACIS observations (2001 October 7)
was taken in CC
mode, which allowed for the first search
of fast pulsations.
An upper limit on pulsed fraction in a 0.03--300~s period range
is $f_p < 15\%$. 

A number of attempts to detect an optical/IR counterpart have been
undertaken. The deepest observations (2001 May-July, with VLT, in
the I,J,H,Ks bands) has shown a very red object at $1\farcs 6$
from the center of the {\sl Chandra} error circle. 
Its IR magnitudes (uncertainties of the values in Table 2 are $\pm$0.5 mag) 
correspond
to M$_{\rm J}$=8.4, M$_{\rm H}$=6.2, M$_{\rm Ks}$=5.4
 at $d=3.3$ kpc, $A_{\rm V}=4.7$. We do not exclude
the possibility that this is a dwarf companion 
or the
long-sought fossil disk (in the former case, the 6-hr period, apparently
seen in the low state, could be a binary period), 
but a more accurate astrometric and
photometric analysis is required
to confirm the identification.

To summarize, the central source of RCW 103 
is the only strongly variable CCO.
The most natural explanation for the variability is accretion
with a variable rate, $\dot{M}\sim 10^{12}$--$10^{14}$ g s$^{-1}$.
The source of accretion, a low-luminosity companion in a binary
system or a dim disk, can be established from further IR observations.
If the IR counterpart is confirmed, this will be the first discovery
of either a very young binary system with a compact companion in a SNR
or a fossil disk which is still present after the SN explosion about
2 kyr ago.
%%%%%%%%%%%%%%%%%%%%%%%%%%%%%%
\subsubsection{Pup A CCO:} This object,
at about $6'$ from the kinematical center of the SNR,
 was discovered with {\sl Einstein}
(Petre et al.\ 1982) and further studied with {\sl ROSAT} and
{\sl ASCA} (Petre, Becker, \& Winkler 1996; Zavlin, Tr\"umper,
\& Pavlov 1999). Large values of $F_{\rm x}/F_{\rm opt}$ (see Table 2)
suggest that this is a NS or a BH.
  The pre-{\sl Chandra} X-ray observations
have shown
that, similar to other CCOs, the spectrum of this object is either a very
steep power law or, more likely, thermal-like, with a BB radius substantially
smaller than $R_{\rm NS}$.
The radius becomes compatible with $R_{\rm NS}$, and the effective
temperature consistent with `standard' NS cooling models
(see Table 4), if one fits
the spectra with a hydrogen atmosphere model with $B>6\times 10^{12}$ G
(Zavlin et al.\ 1999). 
The source was observed 
with {\sl Chandra} ACIS
in 2000 January 1 (12 ks). The spectral parameters obtained from
this observation 
are consistent with those obtained with
{\sl ROSAT} and {\sl ASCA}. 

Observations of this CCO with the {\sl Chandra} HRC
(1999 December 21 and 2001 January 25, 18 ks each) allowed
us to obtain high-resolution images 
and search for pulsations. We found no signs of a PWN around
the source --- an additional confirmation that this is not an active
pulsar. These observations did not confirm a 75-ms
candidate period reported by Pavlov, Zavlin, \& Tr\"umper (1998)
at a $3\sigma$ level. (This result is not surprising because,
with the period and its derivative suggested by Pavlov et al. 1998,
the source should be an active pulsar, in contradiction with
its observed properties.)  Moreover, no other periods are seen in
the power spectra, with an upper limit on pulsed fraction
$f_p < 10\%$ in a 0.003--300 s period range. 
%%%%%%%%%%%%%%%%%%%%%%%%%%%%%%%%%%%%%%%%%%%%%%%%
\subsubsection{CCO in PKS 1209--51/52:} 
The oldest CCO in our sample
was discovered with {\sl Einstein}
(Helfand \& Becker 1984) $6'$ off-center the $81'$ diameter
shell SNR G296.5+10.0. {\sl EXOSAT}, {\sl ROSAT} and {\sl ASCA}
observations (Kellet et al.\ 1987; Mereghetti, Bignami, \& Caraveo
1996; Vasisht et al.\ 1997; Zavlin, Pavlov, \& Tr\"umper 1998)
have established its spectral characteristics similar to those
of the Pup A CCO, albeit with lower temperatures inferred from
thermal fits.
Zavlin et al.\ (1998) have shown
that fitting the CCO spectra with the magnetic hydrogen atmosphere model
yields a radius close to the expected NS radius, an
effective temperature consistent with the standard NS cooling
models, and a hydrogen column density compatible with the values
obtained from independent measurements. 

This object was observed with {\sl Chandra} ACIS
on 2000 January 6 (30 ks) in CC
mode.
Timing analysis allowed us to find a strong ($5\sigma$) candidate
for the period, $P=0.42412924\,{\rm s}\pm0.23\,\mu$s
(Zavlin et al.\ 2000).
The pulsations are smooth, with one peak per period
and pulsed fraction $f_p=9\%\pm2\%$. We can speculate that the
pulsations may be caused by temperature nonuniformity due to
anisotropic heat conduction in a strong magnetic field.
To estimate the magnetic field and elucidate the nature of this
INS, the period derivative must be measured. 

Fits of the {\sl Chandra} spectra
with continuum spectral models give fitting
parameters 
very close to those obtained in the 
pre-{\sl Chandra} observations, but the quality of the
fits is worse now 
because of an apparent broad spectral feature,
presumably caused by inaccuracy of CCD response.
Thus, the {\sl Chandra} ACIS spectrum of this source does not contradict
to the suggestion of Zavlin et al.\ (1998) that this is a NS covered with
a strongly magnetized hydrogen atmosphere.
%%%%%%%%%%%%%%%%%%%%%%%%%%%%%%%%%%%%%%%%%%
\section{Summary}
%%%%%%%%%%%%%%%%%%%%%%%%%%%%%%%%%%%%%%%%%%
Although some properties of CCOs
(in particular, their spectra) are very similar to each other,
this still does not guarantee that they represent a uniform class of objects.
The most outstanding among these sources is the RCW 103 CCO, with
its highly variable flux (time scales hours to months), putative 6-hr period, 
and a possible IR counterpart. We presume that this source is not a
truly isolated NS (i.e., it is powered by accretion, at least in
its high state), although it is not a rotation-powered active pulsar.
The other four CCOs have shown neither long-term variabilities nor
indications of binarity. If we adopt a plausible hypothesis that
their radiation is thermal and assume that it is adequately described
by the same spectral model (e.g., BB or H atmosphere) for each of the sources,
then we have to conclude that the emitting area is growing with
the CCO's age. If we assume that the radiation emerges from a hydrogen
atmosphere, than the size is consistent with a NS radius for the two oldest
CCOs, being smaller for the two younger ones. 
One might suggest that the two younger and two older CCOs
are objects of different types, but it does not answer the fundamental
question: {\it Why are the emitting regions of at least two young CCOs so
small?} If this is due to the above-discussed nonuniformities
of magnetic field or chemical composition, one could speculate
that these nonuformities are formed at very early stages of the NS life,
and they weaken (spread over the NS surface) in a time of several
thousand years.
One could assume that the two younger CCOs are BHs, not
NSs, but then we would have to invoke accretion as an energy source,
which does not look consistent with a lack of variability. 

The key property to understand the nature of CCOs is their periodicity.
So far, the period has been found with high
significance (albeit in only one observation) only for the oldest
of the CCOs, and its value of 424 ms proves that at least this CCO is
a NS. The suspected 12 ms period of the youngest CCO is very
puzzling --- we have to wait for its confirmation before making definitive
conclusions on the nature of this source. 

Another intriguing question is whether the CCOs are close relatives
of AXPs and SGRs, as one could assume based on similarity of their
spectra. Based on the above-mentioned CCO periods
(which, however, require confirmation), the formal answer
should be `no' rather than `yes'.
On the other hand, it is hard to explain why the apparent temperatures and sizes
of emitting regions would be so similar in objects of different
nature. Therefore, we do not rule out the possibility that at least
some of the CCOs belong to the same family as AXPs and/or SGRs,
perhaps at different stages of their evolution. In this respect,
it would be particularly interesting not only to confirm/measure
the CCO periods, but also to evaluate the period derivatives.

To conclude, with the aid of the {\sl Chandra}
observations 
two more CCOs have been discovered,
data of much better quality were obtained, and the main problems
were formulated more clearly.
We expect these problems to be resolved by further observations 
--- X-ray timing
and spectroscopy, supplemented with deep NIR imaging. 
%%%%%%%%%%%%%%%%%%%
\acknowledgements
This work was supported by NASA through grants NAG5-7017, NAG5-10865, NAS8-3852,
and SAO GO0-1012X. We thank A.\ Garmire, O.\ Kargaltsev, and B.\ Kiziltan
for their help with the data analysis.
%%%%%%%%%%%%%%%%%

\end{document}